\documentstyle[twocolumn,prb,aps,epsfig]{revtex}
\begin{document}
\draft

\title{Electrical Transport Properties of Single Crystal ${\text
{Sr}}_3{\text{Ru}}_2{\text {O}}_7$: \\ The Possible Existence of an Antiferromagnetic Instability at Low Temperatures}

\twocolumn[
\hsize\textwidth\columnwidth\hsize\csname@twocolumnfalse\endcsname

\author{Y. Liu, R. Jin, Z. Q. Mao, and K. D. Nelson}
\address{Department of Physics, The Pennsylvania State
University, University Park, PA 16802}
\author{M. K. Haas and R.J. Cava}
\address{ Department of Chemistry and Materials Institute, Princeton
University, Princeton, NJ 08540}

\date{\today}
\maketitle

\begin{abstract}

We report the results of Hall coefficient $R_{\rm H}$ and magnetoresistance (MR) measurements on single crystalline samples of ${\text {Sr}}_3{\text {Ru}}_2{\text
{O}}_7$ grown by the floating zone method. $R_{\rm H}$ was found to be positive over the entire
temperature range studied (0.3 - 300K). Its temperature ($T$)
dependence follows closely that of the magnetic
susceptibility, including a maximum at a characteristic temperature $T$*=17 K.
We show that $R_{\rm H}$ can be decomposed into normal and anomalous parts as in the case of skew scattering in heavy-fermion compounds and ferromagnetic metals.
This, together with the observation that the longitudinal MR
is greater than the transverse MR at the same magnetic field and temperature,
suggests that magnetic fluctuations dominate
the electrical transport properties in ${\text {Sr}}_3{\text {Ru}}_2{\text
{O}}_7$. We found a crossover in the sign of the MR at
$T$*, from positive to negative as the temperature increased, for both the
transverse and the longitudinal configurations. In addition, a non-monotonic behavior in the field dependence of the MR was found at low temperatures.  These observations
suggest that the magnetic correlations in ${\text {Sr}}_3{\text {Ru}}_2{\text {O}}_7$ at ambient pressure undergo a qualitative change as the temperature is lowered. Above $T$*, they are dominated by ferromagnetic instability. However, below $T$*, the system crosses over to a different behavior, controlled possibly by a canted antiferromagnetic instability.

\end{abstract}

\pacs{74.70.-b, 74.25.Fy, 74.25.Ha, 74.72.Yg}

]


\narrowtext

\section {INTRODUCTION}

The discovery of superconductivity in the layered perovskite
Sr$_2$RuO$_4$ \cite{1}, which appears to posses a
$p$-wave spin-triplet pairing \cite{2,3,4}, has renewed interest in the magnetic properties of related compounds, such as
SrRuO$_3$ and Sr$_3$Ru$_2$O$_7$. Understanding the magnetic
properties of these compounds may provide clues to the mechanism
leading to $p$-wave pairing in Sr$_2$RuO$_4$.

SrRuO$_3$, the three-dimensional (3D), cubic perovskite in
the Ruddlesden-Popper (R-P) series (Sr$_{n+1}$Ru$_n$O$_{3n+1}$,
with $n=\infty $), is an established ferromagnet with $T_c$ = 160 K
\cite{5,6}. For Sr$_3$Ru$_2$O$_7$, the n=2 member in the R-P series, has been the subject of controversy regarding its magnetic properties \cite{7,8,9,10,11}.  In the original work of Cava {\it et al} on phase-pure, polycrystalline ${\text {Sr}}_3{\text {Ru}}_2{\text {O}}_7$ \cite{7}, the magnetic susceptibility $\chi $ was found to show a peak around 15 K, accompanied by a Curie-Weiss behavior at high temperatures, $\chi=\text {C}/(T-{\theta}_{\text {CW}})$, where C is the Curie constant, $T$ is the temperature, and $\theta_{\text {CW}}$ is the Curie temperature, yielding negative ${ \theta}_{\text{CW}}\approx$ 15 K and a large moment of 2.5 $\mu_{\text {B}}$/Ru. 

In a subsequent study, using single crystals of Sr$_3$Ru$_2$O$_7$
prepared by flux method,
Cao {\it et al}  \cite{8} reported a ferromagnetic (FM) ordering at 104 K
under ambient pressure. This result was in sharp contrast with the neutron diffraction results on polycrystalline
${\text {Sr}}_3{\text {Ru}}_2{\text {O}}_7$, which did not reveal any
long-range magnetic ordering down to 1.6 K  \cite{9}. Recently,
Ikeda {\it et al}. \cite{10} reported the magnetic susceptibility
($\chi (T)$) data of Sr$_3$Ru$_2$O$_7$ obtained on crystals
prepared by the FZ method, which shows a peak at about 17 K but no
ferromagnetic ordering. This result is different from Cao's result, but consistent with earlier results obtained in polycrystals. It has been
suggested that the FM ordering observed in flux crystals was induced by contamination from impurities of the flux, the crucible, or a combination of both \cite{10}.

Ikeda {\it et al}. also studied the magnetic properties
of Sr$_3$Ru$_2$O$_7$ under hydrostatic pressure. Clear evidence for FM
ordering was found. The FM transition temperature, $T_c$, was found to
be 70 K at 1.1 GPa.  Therefore the authors argued that
Sr$_3$Ru$_2$O$_7$ is a nearly FM Fermi liquid. They furtherfore suggested that the
maximum behavior in $\chi (T)$ might be due to the critical spin fluctuations found close to a quantum critical point, as in the case of
(Ca,Sr)$_2$RuO$_4$ \cite{12} and MnSi \cite{13}.

To further clarify the nature of the magnetic fluctuations
in Sr$_3$Ru$_2$O$_7$, we have carried out
Hall coefficient ($R_{\rm H}$) and magnetoresistance
(MR) measurements using Sr$_3$Ru$_2$O$_7$ single crystals
grown by the FZ method. In this article, we
will present the results of our measurements and discuss
their physical implications. 

\section {EXPERIMENTAL METHODS}

Single crystals of ${\text {Sr}}_3{\text {Ru}}_2{\text {O}}_7$ used in
this study were grown by the floating-zone method. $X$-ray diffraction
measurements confirmed a crystal structure of 

\begin{figure}
\centerline{\epsfig{file=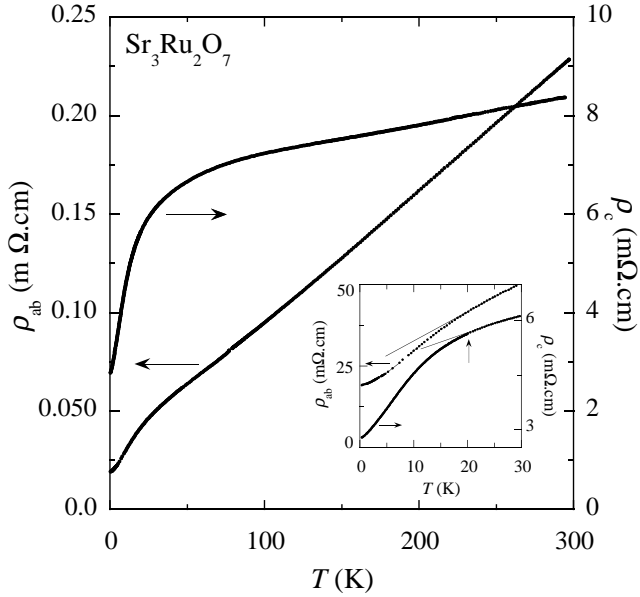,angle=0,width=3.4in}}
\medskip 
\caption{Temperature dependence of resistivity of
Sr$_3$Ru$_2$O$_7$ single crystal. The inset shows that both
$\rho_{ab}$ and $\rho_c$ have a slope change around 17 K.}
\label{1}
\end{figure}

\noindent the $n$=2 R-P compound, showing no impurity phases. For in-plane MR and Hall measurements, we used two rectangular-shaped single crystals with dimensions around $0.5\times 0.1\times 0.1$ and $0.8\times 0.3\times 0.15 {\text {mm}}^3$, respectively. For each sample, two current contacts covering the opposite ends and four voltage contacts on the two
sides of the crystal were prepared.  All ${\text {RuO}}_2$ layers were
electrically shorted along the {\it c}-axis to ensure a homogeneous current
distribution. The {\it c}-axis transverse ($H\perp I$) MR measurements were
carried out in a single crystal with dimensions around $0.8\times 0.4\times
0.2 $ mm$^3$. Two ring-shaped current contacts were prepared on the
opposite $ab$ faces. The two voltage contacts were point-like positioned in
the center of the rings.

Electrical measurements were carried out in a $^3$He and
dilution refrigerator. The temperature was measured using a Lakeshore
Cernox 1030 thermometer with relative temperature corrections (due to the applied magnetic field, typically 0.15\% at 4.2 K and 5.9\% at 2K and 8.0T). For
transverse and longitudinal MR measurements, the magnetic
field $H$ was applied perpendicular and parallel to the injected current
$I$, respectively. In order to exclude the Hall contribution to the MR, only
the symmetric part of $\Delta \rho _{ab}(H)=\rho _{ab}(H) - \rho _{ab}(0)$
under field reversal was included. For Hall measurements, the magnetic
field was applied parallel to the $c$-axis with a current bias applied along the
$ab$-plane. The Hall voltage $V_{\rm H}$, which contains only the
asymmetric contributions under field reversal, was found to vary
linearly with $H$ up to 4 T over the whole temperature region. By fitting $V_{\rm H}(H)$ data using $V_{\rm
H}=R_{\rm H}\cdot H\cdot I/d$ ($d$ is the thickness of the sample along the
$c$-axis), the Hall coefficient $R_{\rm H}$ was obtained.

\begin{figure}
\centerline{\epsfig{file=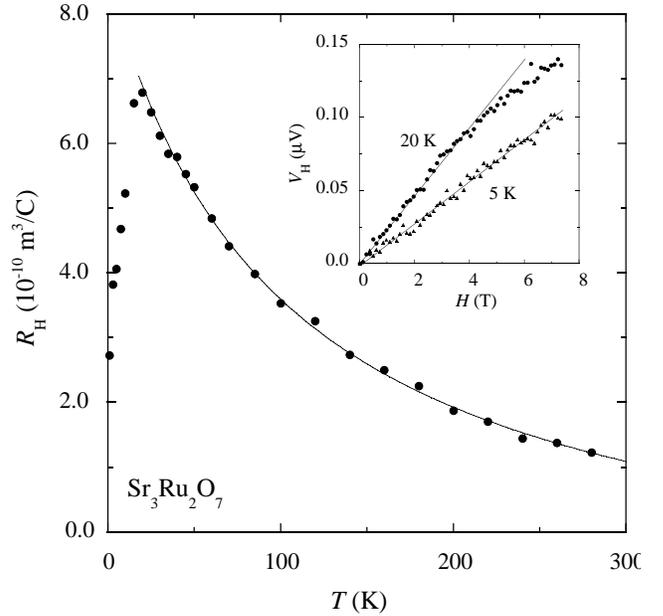,angle=0,width=3.4in}}
\medskip 
\caption{In-plane Hall coefficient $R_{\text
H}(T)$ for ${\text {Sr}}_3{\text {Ru}}_2{\text {O}}_7$ with a
peak around 17K.  The inset shows the magnetic field dependence
of $V_{\text H}$ at two different temperatures.}
\label{2}
\end{figure}

\section {EXPERIMENTAL RESULTS AND DISCUSSIONS}

Figure \ref{1} shows the temperature dependence of electric
resistivity for in- and out-of-plane directions of the
Sr$_3$Ru$_2$O$_7$ crystals (denoted by $\rho_{ab}$ and $\rho_c$
respectively). The overall shape of these data are similar to
those reported in Ref.10. Both $\rho_{ab}(T)$ and
$\rho_c(T)$ are metallic in the whole temperature region. The
sharp drop in $\rho_c (T)$ below 50 K has been attributed to the suppression of thermal scattering between quasi-particles and phonons \cite{10}. Interestingly, both $\rho_{ab}$ and $\rho_c$ were found to show an apparent slope change around 17 K (see the inset of Fig. \ref{1}). This temperature is close to that below which the magnetic susceptibility drops sharply, indicating that magnetic fluctuations have strong influence on the electrical transport properties.

The temperature dependence of the Hall coefficient $R_{\rm H}$ is
shown in Fig. \ref{2}. $R_{\rm H}$ is positive over the
entire temperature region studied (0.30-300K), reaching a maximum
at the same temperature (=17 K) where both $\rho_{ab}(T)$ and
$\rho_c(T)$ show a slope change ($T$* will be used to denote
this characteristic temperature). These observations agree
well with a previous report \cite{14} in which $R_{\rm H}$
was measured on single crystals prepared by the FZ method as well.

The temperature dependence of $R_{\rm H}(T)$ is strikingly similar to that
of $\chi(T)$ \cite{7,10}. Such a behavior was observed previously in
heavy fermion compounds such as in UPt$_3$ \cite{15} and ferromagnetic
metals \cite{16}.  It was found that $R_{\rm H}$ can be fit by
        \begin{equation}
        $$R_{\text H} = R_{\text 0}+ R_{\text s}\times 4\pi\chi(T)$$
        \eqnum{1}.
        \end{equation}

\begin{figure}
\centerline{\epsfig{file=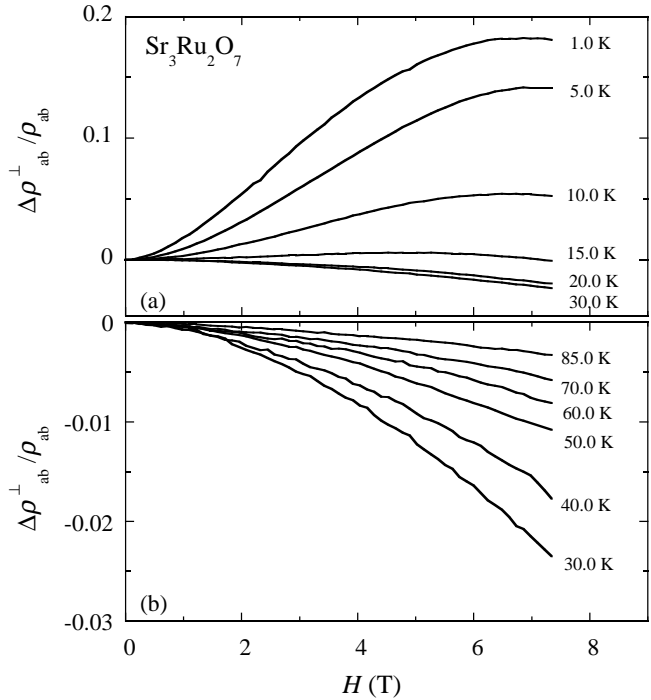,angle=0,width=3.4in}}
\medskip 
\caption{In-plane transverse MR, $\Delta \rho
_{ab}^{\perp }/\rho _{ab}$ ($H\perp {ab}$, $I\parallel ab$), for
${\text {Sr}}_3{\text {Ru}}_2{\text {O}}_7$.  A sign change in MR
is seen at 17K.}
\label{3}
\end{figure}

\noindent $R_{\text 0}$ and $R_{\text s}$ are
referred to as the normal and anomalous Hall coefficients
respectively. Such behavior of
$R_{\rm S}$ can be understood in the picture of a skew scattering
 that involves competition between spin-orbit coupling and spin-flip
scattering \cite{17}.

For Sr$_3$Ru$_2$O$_7$, it has been shown that $\chi (T)$
satisfies the Curie-Wiess law \cite{7,10} at high temperatures. We
have attempted to fit our $R_{\rm H}$ data shown in Fig.\ref{2} using
the following equation,
       \begin{equation}
       $$R_{\text H} = R_{\text 0}+ R'_{\text s}/
(T - \theta _{\text {CW}})$$
       \eqnum{2},
       \end{equation}
with three fitting parameters. The best fit (see the solid line in Fig. \ref{2}) was obtained at $R_{\rm 0} = -1.4\times 10^{-10}$ m$^3$/C,
$\theta_{\rm {CW}}$ = -102 K, and $R'_{\rm S}$ = 1.7$\times
10^{-7}$ m$^3$/C. The absolute value of $\theta_{\rm {CW}}$ so
obtained is higher than that obtained by fitting $\chi (T)$
curves (around -15 K for polycrystals and -40 K for single
crystals), but has the same sign as that found from magnetic
susceptibility measurements. $R_{\rm 0}$ was found to be negative, which
means that the normal Hall effect is dominated by electrons. This is
consistent with our expectation since band structure calculations reveal
that in Sr$_3$Ru$_2$O$_7$ there are four electron- and two
hole-like bands crossing the Fermi surface \cite{18}. These
observations suggest that skew scattering is a reasonable
picture for understanding our $R_{\rm H}(T)$ result.

\begin{figure}
\centerline{\epsfig{file=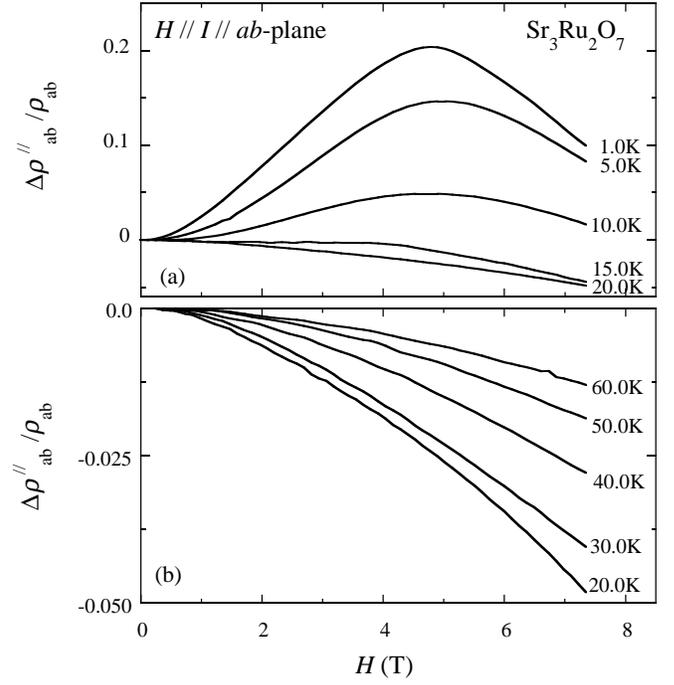,angle=0,width=3.4in}}
\medskip 
\caption{In-plane longitudinal MR $\Delta \rho
_{ab}^{\parallel }/\rho _{ab}$ ($H\parallel {ab}$, $I\parallel ab$)
for ${\text {Sr}}_3{\text {Ru}}_2{\text {O}}_7$.}
\label{4}
\end{figure}

As pointed out in Ref. 19, a negative $\theta_{\rm {CW}}$
does not mean that the magnetic fluctuations are necessarily
anti-ferromagnetic (AFM) in nature, as frequently assumed in literature. In fact,
we found that $V_{\rm H}$ as a function of magnetic field deviated from
linear behavior above $T$* (see inset of Fig. 2) in a manner characteristic
of ferromagnetic behavior, consistent with the argument of nearly FM behavior in Ref.10. However, below $T$*, $V_{\rm H}$ becomes linear
with $H$.  

Figure \ref{3} shows the transverse in-plane MR, $\Delta
\rho_{ab}^{\perp }/\rho_ {ab}$ ($H\perp I $), between 0 and 7.3 T
at various temperatures. At high temperatures, $\Delta \rho
_{ab}^{\perp }/\rho _{ab}$ is negative and small, varying
monotonically with $H$. The magnitude of $\Delta \rho
_{ab}^{\perp }/\rho _{ab}$ increases with decreasing temperature
down to 30 K. As $T$ approaches $T$* from high temperatures,
the opposite tendency is observed. The value of $\Delta \rho _{ab}^{\perp
}/\rho _{ab}$ eventually becomes positive with $T<T$* (=17 K).
Another striking feature is that, below
$T$*, $\Delta \rho _{ab}^{\perp }/\rho _{ab}$
shows a non-monotonic field-dependence. With increasing $H$,
$\Delta \rho _{ab}^{\perp }/\rho _{ab}$ initially increases then
decreases after reaching a maximum at $H_{\rm {max}}$, which
appears to increase with decreasing temperature. This
non-monotonic behavior of $\Delta \rho_{ab}^{\perp
}/\rho_{ab}$($H$) persists to the lowest temperature measured.

We also measured the longitudinal in-plane MR, $\Delta
\rho_{ab}^{\parallel }/\rho_{ab}$, of Sr$_3$Ru$_2$O$_7$. As shown
in Fig. \ref{4}, $\Delta \rho _{ab}^{\parallel }/\rho _{ab}$
exhibits similar features as $\Delta \rho _{ab}^{\perp }/\rho
_{ab}$. In particular, it also shows a sign reversal at 
temperatures slightly lower than $T$*. A more pronounced
non-monotonic field-dependence was also found for $\Delta \rho
_{ab}^{\parallel }/\rho _{ab}$ for $T<$15 K.

\begin{figure}
\centerline{\epsfig{file=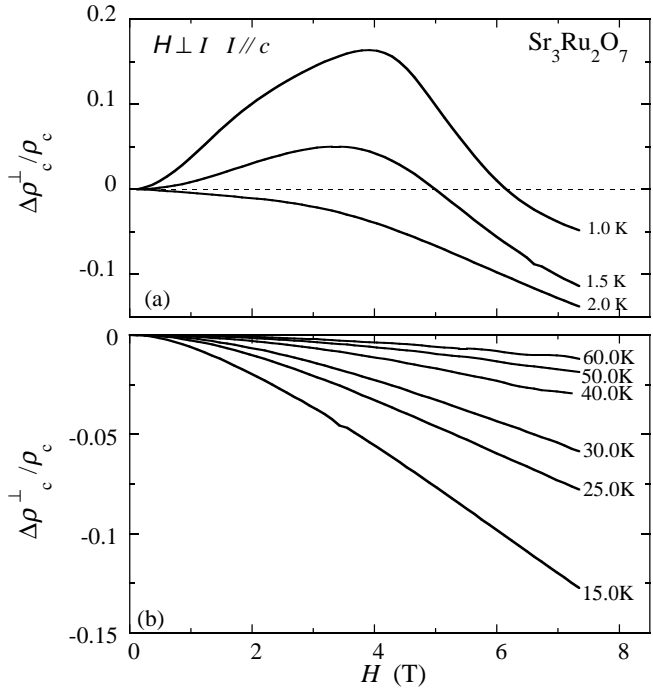,angle=0,width=3.4in}}
\medskip 
\caption{c-axis transverse MR,
$\Delta \rho _{c}^{\perp }/\rho _{c}$ ($H\perp {c}$, $I\parallel
c$), for ${\text {Sr}}_3{\text {Ru}}_2{\text {O}}_7$.}
\label{5}
\end{figure}

We would like to point out another important
feature in Fig. \ref{3} and \ref{4}: The magnitude of $\Delta \rho
_{ab}^{\parallel}/\rho _{ab}$ is greater than that of $\Delta
\rho _{ab}^{\perp }/\rho_{ab}$ at the same magnetic field
below approximately 10 K. Since the current is not subject to the Lorentz force in the longitudinal configuration, it reflects mostly the contribution of spins. Therefore, this observation, together with that of the skew scattering, suggests that spin scattering has a strong influence on the electrical transport in Sr$_3$Ru$_2$O$_7$.

Figure \ref{5} shows the transverse $c$-axis MR ($\Delta
\rho_{c}^{\perp }/\rho_{c}$) of Sr$_3$Ru$_2$O$_7$ at different
temperatures. It is clear that $\Delta \rho_{c}^{\perp
}/\rho_{c}$ exhibits all the features observed in the in-plane MR.
The only obvious difference is that the negative-to-positive sign
reversal in $\Delta \rho_{c}^{\perp }/\rho_{c}$ was found at a
slightly lower temperature(10 K).

It is remarkable that a maximum in $\chi (T)$ and $R_{\rm H}(T)$,
a slope change in $\rho_{ab}(T)$ and $\rho_c(T)$, a deviation from linear magnetic field dependence in $V_{\rm H}$, and a sign change
in longitudinal and transverse MR, were all
found at approximately the same temperature. These observations suggest that in Sr$_3$Ru$_2$O$_7$, instead of continuing its trend to move closer to FM ordering as temperature is lowered, as one would naturally expect, the system is side tracked to a different behavior below $T$*. This qualitative change in the dynamics of the system has to be magnetic in origin.

\mbox{Physical insight may be obtained from the} (Ca$_{2-x}$Sr$_x$)RuO$_4$ solid 
solution system \cite{12}. For $x < 0.2$, (Ca$_{2-x}$Sr$_x$)RuO$_4$ 
is antiferromagnetic.  For $0.2\le x\le 0.5$, this material system is near a FM 
instability. In particular, at $x\simeq 0.5$, it is nearly ferromagnetic, 

\begin{figure}
\centerline{\epsfig{file=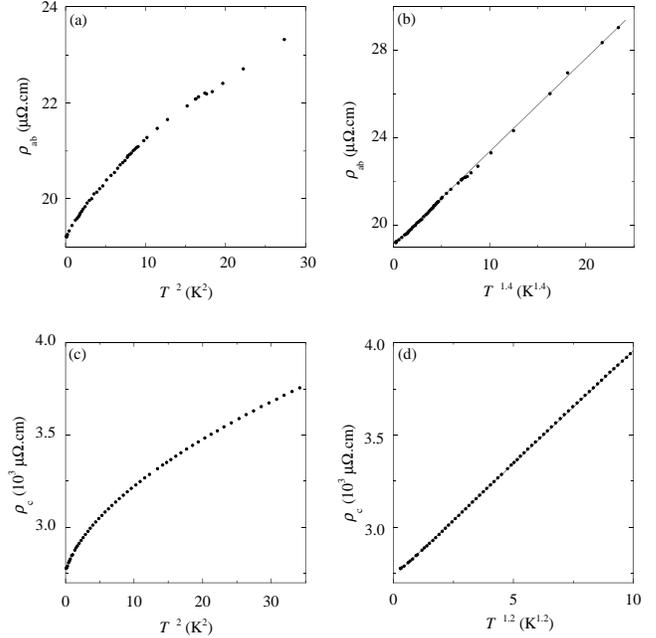,angle=0,width=3.4in}}
\medskip 
\caption{(a) and (c) show the deviation of $\rho _{ab}$ and $\rho _c$ from $T^2$ dependence. The best fits obtained are $T^{1.4}$ for $\rho _{ab}$ (b) and $T^{1.2}$ for $\rho _c$ (d).}
\label{6}
\end{figure}

\noindent evolving from paramagnetic Sr$_2$RuO$_4$ through
band narrowing.  Between $0.2\le x < 0.5$, it changes into a
state with short-range AFM ordering below a
characteristic temperature $T_{\rm P}$ ($T_{\rm P}$, which is about 10 K
at $x=0.2$, decreases continuously to zero at $x=0.5$). Below $T_{\rm P}$,
$\rho_{ab}(T)$ shows a change of slope.  MR shows a negative-positive
sign reversal and non-monotonic behavior in field dependence.  In addition,
 both $\chi (T)$ and $R_{\rm H}$ show a maximum at $T_{\rm P}$.

The phenomena seen in (Ca$_{2-x}$Sr$_x$)RuO$_4$ are clearly similar to those of Sr$_3$Ru$_2$O$_7$. Furthermore, (Ca$_{2-x}$Sr$_x$)RuO$_4$ and Sr$_3$Ru$_2$O$_7$ have comparable electronic specific heat coefficients and Wilson ratios. All these similarities suggest that the magnetic correlations in Sr$_3$Ru$_2$O$_7$ show a qualitative change below $T$*. At high temperatures, FM spin fluctuations dominate because of the close proximity to the FM instability, while at low temperatures (below $T$*) they cross over to AFM fluctuations.

In the magnetization ($M$) vs. field ($H$) curve obtained on
Sr$_3$Ru$_2$O$_7$ poly-crystals \cite{7,11}, $M(H)$
shows a convex character at low fields below
$T$*.  This is a characteristic feature of AFM correlations.
Above $T$*, $M(H)$ shows a concave feature,
typical of FM correlations. This further supports the observation that as the temperature decreases, Sr$_3$Ru$_2$O$_7$ at ambient pressure moves away from an FM, but closer to an AFM magnetic instability. This scenario seems to be capable of explaining all features seen in the electrical transport in
Sr$_3$Ru$_2$O$_7$ described above.

In (Ca$_{2-x}$Sr$_x$)RuO$_4$, the change of magnetic correlations
is driven by a structural phase transition at temperature
above $T_{\rm P}$ \cite{19}. In Sr$_3$Ru$_2$O$_7$, although
neutron diffraction measurements did not reveal any structural
transition \cite{9}, an unusual change in structural details at
low temperature, i.e., a negative thermal expansion along the
$c$-axis, was observed. This might be responsible for the change of magnetic coupling at low temperatures.

We note, however, that the anisotropy in $\chi (T)$
between the in- and out-of-plane directions in
Sr$_3$Ru$_2$O$_7$ single crystals \cite{10} is quite small.
In addition, $\chi (T)$ was found to saturate below roughly 6 K.
How do we explain these observations in the picture of AFM fluctuations below $T$*? 

This apparent difficulty may be solved if we
assume that the spins aligned anti-ferromagnetically in
Sr$_3$Ru$_2$O$_7$ are canted to the $c$-axis. A small net ferromagnetic
component in the c-axis in the canted AFM can explain the
slight hysteresis observed in the $M-H$ curve on poly-crystals Sr$_3$Ru$_2$O$_7$ \cite{11}. The saturation of $\chi (T)$ below 6 K is likely to imply that the crossover from FM to AFM is incomplete, probably because Sr$_3$Ru$_2$O$_7$ does not undergo a structural phase transition as in (Ca$_{2-x}$Sr$_x$)RuO$_4$. That might also be the reason why neutron diffraction did not detect any sizable magnetic ordering. 

For (Ca$_{2-x}$Sr$_x$)RuO$_4$, the magnetic instability in
the $0.2\le x\le 0.5$ region leads to  non-Fermi liquid
behavior. Its $\rho_{ab}(T)$ shows a $T^{1.4}$ dependence. For
Sr$_3$Ru$_2$O$_7$, our resistivity data exhibit a similar
behavior. Both $\rho_{ab}(T)$ and $\rho_c(T)$ clearly
deviate from $T$-squared dependence. The best fit is $T^{1.4}$ for $\rho_{ab}(T)$ and $T^{1.2}$ for $\rho_c(T)$, as shown in
Fig. \ref{6}. However, we note that the residual resistivity $\rho_0$
estimated from Fig. \ref{1} is about 19.0 (2.8) $\mu \Omega \cdot $cm for
$\rho_{ab}$ ($\rho_c$). It is about five times (for
$\rho_{ab}$) and two times (for $\rho_c$) greater than that reported by
Ikeda in Ref.10, suggesting that our FZ crystals may contain slightly more impurities and defects than theirs.  As a result, it is natural to ask whether the observed non-$T^2$ behavior is intrinsic to Sr$_3$Ru$_2$O$_7$. In Ref.10, it was argued that Sr$_3$Ru$_2$O$_7$ is a Fermi liquid with
$\rho_{ab}(T)\propto T^2$. However, their $T$-squared fitting for
resistivities, especially for $\rho_{ab}(T)$, appears to be less than the best fit. A deviation from the $T$-squared dependence in $\rho_{ab}(T)$ 
is clearly visible even below 6 K. This deviation suggests an intrinsic deviation from Fermi liquid behavior in Sr$_3$Ru$_2$O$_7$, most probably due to the system being close to an AFM instability.

The non-monotonic field-dependence of MR observed in
Sr$_3$Ru$_2$O$_7$ below $T$* can be easily understood in the crossover scenario as discussed above, in which the MR should contain both negative and positive terms. The negative term must have originated from the development of FM spin fluctuations as the field increases.

\section {CONCLUSION}

In summary, we have studied the transport properties of Sr$_3$Ru$_2$O$_7$ under magnetic field using single crystals prepared by the FZ
method. The temperature dependence of the in-plane Hall
coefficient has been found to resemble that of the magnetic
susceptibility, showing a maximum at $T$*. The field dependence of $V_{\rm H}$ was found to deviate from linear behavior above $T$*. Both
$\rho_{ab}(T)$ and $\rho_c(T)$ exhibit a slope change at about $T$*. Furthermore, we found that the longitudinal
in-plane MR is larger than the transverse. Both the in-plane and $c$-axis MR show a negative-to-positive sign reversal below $T$*. All these observations support our assessment that the magnetic correlations in Sr$_3$Ru$_2$O$_7$ are dominated by FM fluctuations above $T$*, but cross over to AFM behavior blow $T$*. Finally, the ground state may deviate from the conventional Fermi-liquid behavior.

We acknowlege useful discussions With Prof. Y. Maeno and Dr. S-I
Ikeda. Y. Liu has benefited from communications with Dr. G. Cao.
This work is supported by NSF through grants DMR-9702661 and
ECS-9705839 at Penn State and DMR-9808941 at Princeton.



\end{document}